\documentclass[APISAT]{tjsass} 

\usepackage[dvipdfmx]{graphicx}
\RequirePackage{multicol}
\RequirePackage{amsmath}
\RequirePackage[varg]{txfonts}
\RequirePackage{bm}
\RequirePackage{array}
\usepackage{color}
\RequirePackage{tjsasscite}
\usepackage{dblfloatfix}

\title{Vision-based Target Pose Estimation with Multiple Markers \\
for the Perching of UAVs}
\subtitle{}
\author{\NAME{Truong-\uppercase{D}ong}{DO}\thanksNum{1,3)}, 
\NAME{Nguyen}{XUAN-MUNG}\thanksNum{2)} and
\NAME{Sung-\uppercase{K}yung}{HONG}\thanksNum{1,2)}\CorresAuthor{skhong@sejong.ac.kr}}
\thanksOrg{Department of Convergence Engineering for Intelligent Drone, Sejong University, Seoul, South Korea}
\thanksOrg{Faculty of Mechanical and Aerospace Engineering, Sejong University, Seoul, South Korea}
\thanksOrg{Department of Aerospace Engineering, Sejong University, Seoul, South Korea}
\begin{abstract}
Autonomous Nano Aerial Vehicles have been increasingly popular in surveillance and monitoring operations due to their efficiency and maneuverability. Once a target location has been reached, drones do not have to remain active during the mission. It is possible for the vehicle to perch and stop its motors in such situations to conserve energy, as well as maintain a static position in unfavorable flying conditions. In the perching target estimation phase, the steady and accuracy of a visual camera with markers is a significant challenge. It is rapidly detectable from afar when using a large marker, but when the drone approaches, it quickly disappears as out of camera view. In this paper, a vision-based target poses estimation method using multiple markers is proposed to deal with the above-mentioned problems. First, a perching target with a small marker inside a larger one is designed to improve detection capability at wide and close ranges. Second, the relative poses of the flying vehicle are calculated from detected markers using a monocular camera. Next, a Kalman filter is applied to provide a more stable and reliable pose estimation, especially when the measurement data is missing due to unexpected reasons. Finally, we introduced an algorithm for merging the poses data from multi markers. The poses are then sent to the position controller to align the drone and the marker’s center and steer it to perch on the target. The experimental results demonstrated the effectiveness and feasibility of the adopted approach. The drone can perch successfully onto the center of the markers with the attached 25mm-diameter rounded magnet.
\end{abstract}
\begin{document}
\maketitle






\section{Introduction}

Unmanned Aerial Vehicles (UAVs), such as drones and multicopters, have become a popular research topic because of their high autonomy and maneuverability \cite{b1,b2,b3,b4}. With the growing popularity of UAVs, it is necessary to increase awareness of the environment while improving flight performance.The small size and light weight of nano UAVs make them ideal platforms \cite{b5,b6,b7,b8}. However, they have a very limited flight time. Fortunately, many missions do not require hovering for the entire duration. Therefore, it is necessary to develop autonomous perching solutions to conserve energy.

Perching refers to supporting the aerial robot's weight from within using grasping, attachment, or embedding techniques \cite{b9}. This capability could also be useful for tasks requiring the robot to maintain a precise, static position, act as a radio relay in disaster zones, or suspend operation during unfavorable weather. Thus, perching is an appealing ability for aerial vehicles.

The advancement of embedded systems has resulted in a variety of small cameras that can be mounted on UAVs \cite{b10}. These camera modules provide simple, affordable, and reliable solutions that can greatly improve navigation systems \cite{b11,b12}. In the field of visual servoing, there is foundational literature covering control using monocular vision, which discusses the differences between Position Based Visual Servoing (PBVS) and Image-Based Visual Servoing (IBVS) \cite{b13,b14}. The visual servoing approaches \cite{b15,b16} have shown autonomous perching results without the use of motion capture but are highly dependent on objects’ shapes and require the object to initially be in the field of view. Cho et al.\cite{b17} proposed an algorithm to estimate the pose of the robot based on a set of customized markers located on the docking object.

\begin{figure}[b!]
\centering
\includegraphics[width=7cm]{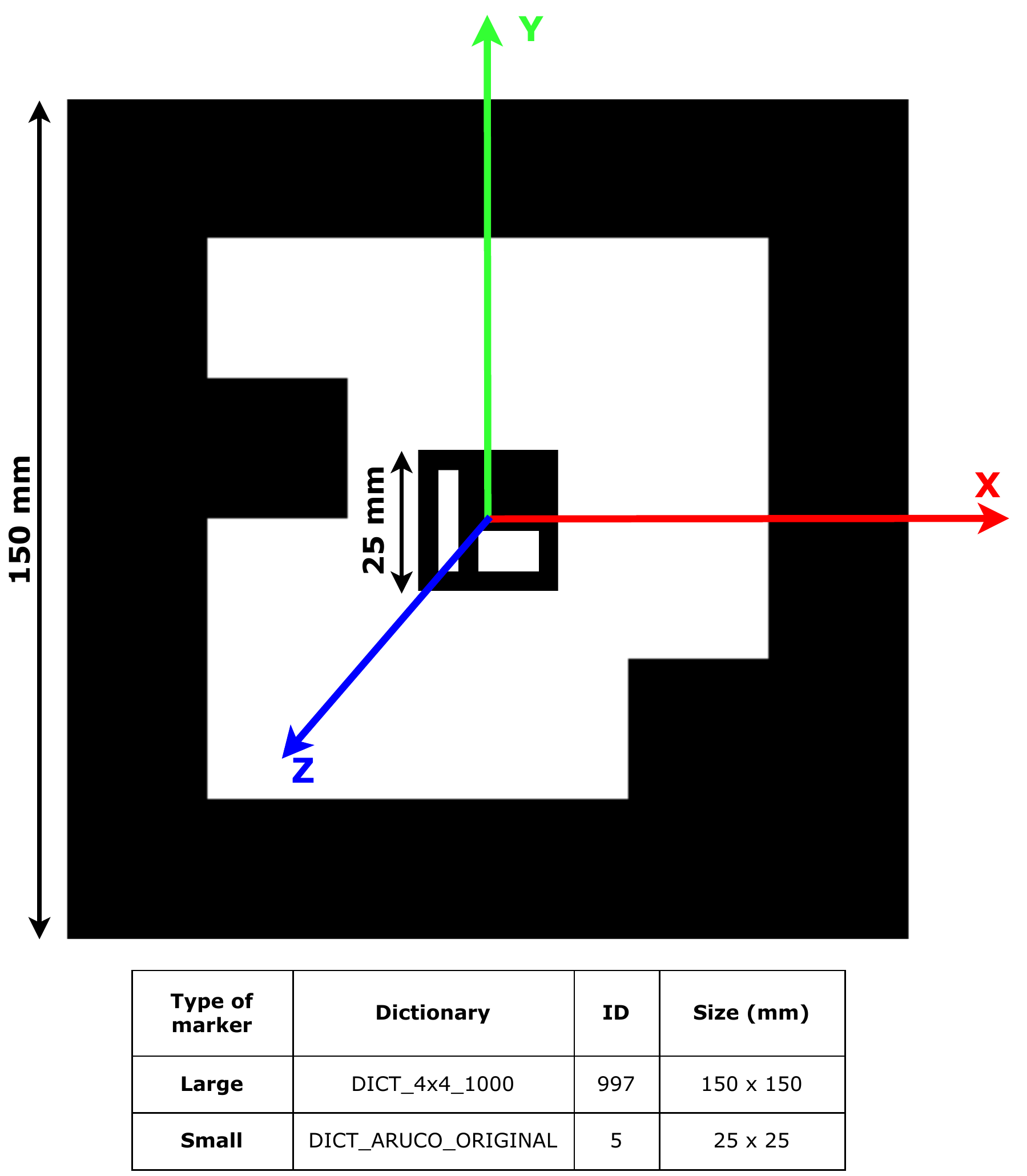}
\caption{The designed multi-marker perching target.}
\centering
\label{fig:multi-marker}
\end{figure}

\begin{figure*}[t!]
\centering
\includegraphics[width=0.8\textwidth]{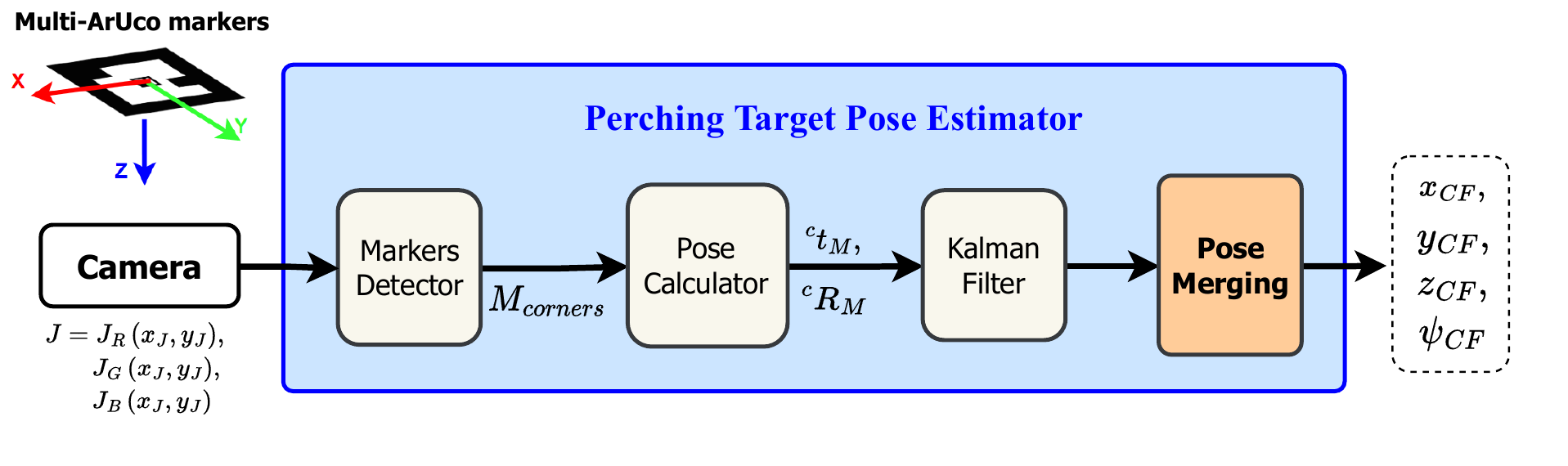}
\caption{Block diagram of Perching Target Pose Estimator using multiple markers}
\label{fig:proposed_estimator}
\end{figure*}

\begin{figure*}[b!]
\centering
\includegraphics[width=0.6\textwidth]{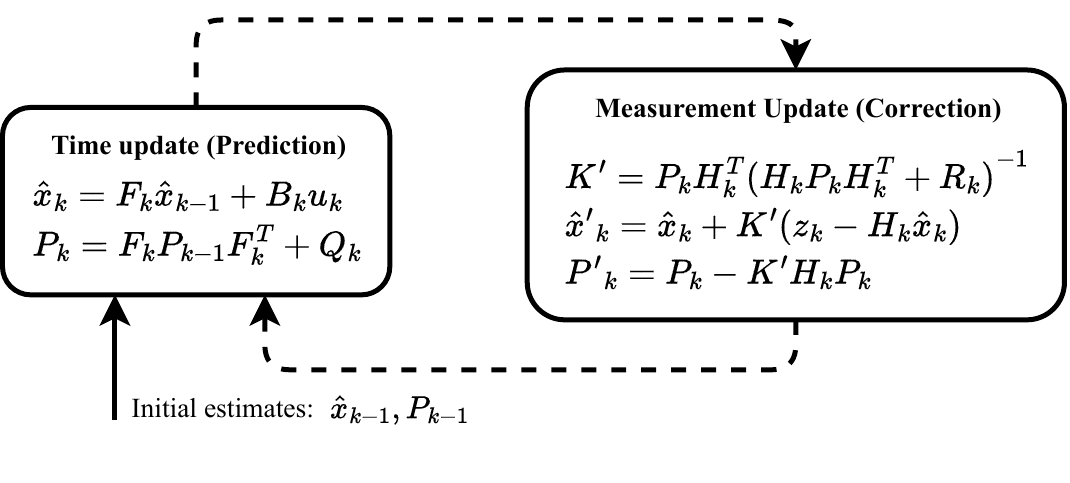}
\caption{Operation of Kalman Filter.}
\label{fig:kalman}
\end{figure*}

ArUco \cite{b18} marker is a binary fiducial marker that composes an internal binary matrix and a black boundary. Large markers are good for the distant detection of the camera. However, when the sensor and the markers are close enough, the markers can be out of the sensor’s field of view that causing the target loss phenomenon and the perching failure. In contrast, a small marker offers the advantage of being detectable when the drone approaches, but it is difficult to identify from a distance.




In this work, we use a camera to detect multiple markers (ArUco) and then merge them to estimate poses. First, the perching target is designed to enhance detection capability at both a wide and close distance. Second, the pose of these markers was estimated using the image gathered by the vision camera mounted on the vehicle. Third, a Kalman filter is applied to provide a more accurate and consistent pose, especially when measurement data is missing due to transmission losses, camera vibrations, or missing detection data. Next, we introduced an algorithm for fusing the poses from calculated data. Finally, we build a testbed for real experiments to validate the strategy and allow the performance of the adopted techniques to be evaluated. The preliminary results indicate the viability of our proposal for autonomous nano-UAVs perching research.

The remainder of the paper is organized as follows. Section 2 provides the methodologies used to accomplish this task. The experimental results obtained are demonstrated in Section 3. Then, conclusions are presented in Section 4, along with future research directions.

\section{Methodology}

The perching target, including a small marker located inside a large one, is described in Fig. \ref{fig:multi-marker}. A larger maker's size will improve detection accuracy but will also increase computational complexity and processing time. Thus, we chose simple markers to detect in accordance with the requirement for quickness. The larger square marker has a size of \(150mm\) and belongs to the $DICT\_4x4\_100$ dictionary with the $ID = 997$. Meanwhile, to prevent the detect confusion, the smaller marker is chosen from $DICT\_ARUCO\_ORIGINAL$ with $ID = 5$. It is placed at the center of the larger marker with $25mm$ in the dimension. Both have the same origin and coordinate; the inclusion of the smaller marker may harm the detection of the larger one. Nonetheless, in our tests, this problem did not occur.

Figure \ref{fig:proposed_estimator} demonstrates the block diagram of the proposed multi-markers based perching target pose estimation algorithm. Before we can begin using the ArUco library for marker detection and pose estimation, the camera must be calibrated to be able to recover information about the size of the markers in real-world units and determine the pose of the camera in the scene with good accuracy. The ArUco module provides this option through OpenCV \cite{b19} from which we can obtain the camera intrinsic parameters and the distortion coefficients.

Through the ArUco library from OpenCV, each obtained image frame $(J)$ is processed in the computer and once the markers $({{M}_{corners}})$ has been detected by the algorithm, the pose of the quadcopter relative to each marker $({}^{c}{{R}_{M}},{}^{c}{{t}_{M}})$ is calculated by Perspective-n-Point $(Pnp)$ solver algorithm \cite{b20} and is fed to the Kalman filter \cite{b21}. After merging them, a precise relative pose is achieved so that Crazyflie can align itself towards the perching target. The output of the proposed pose estimator is comprised of the relative heading angle $(\psi_{CF})$ and translation in $x$, $y$, and $z$-axis $(x_{CF}, y_{CF}, z_{CF})$. All the estimated pose values between the target and the drone are in the world frame centimeters.  

\subsection{Noise and incomplete data filtering}
Because of the noise introduced by the limited capabilities of the camera, transmission losses, or loss of detection, it requires a method for estimating the relative pose of the drone given the incomplete or noisy data from the images. The Kalman filter cleans up the measured data and projects the measurement onto a state estimate. It addresses the problem of predicting the state of a dynamical system at a discrete-time step $k$, given measurements from the current state at the time step $k-1$ and its uncertainty matrix.
\begin{figure*}[b!]
\centering
\includegraphics[width=\textwidth]{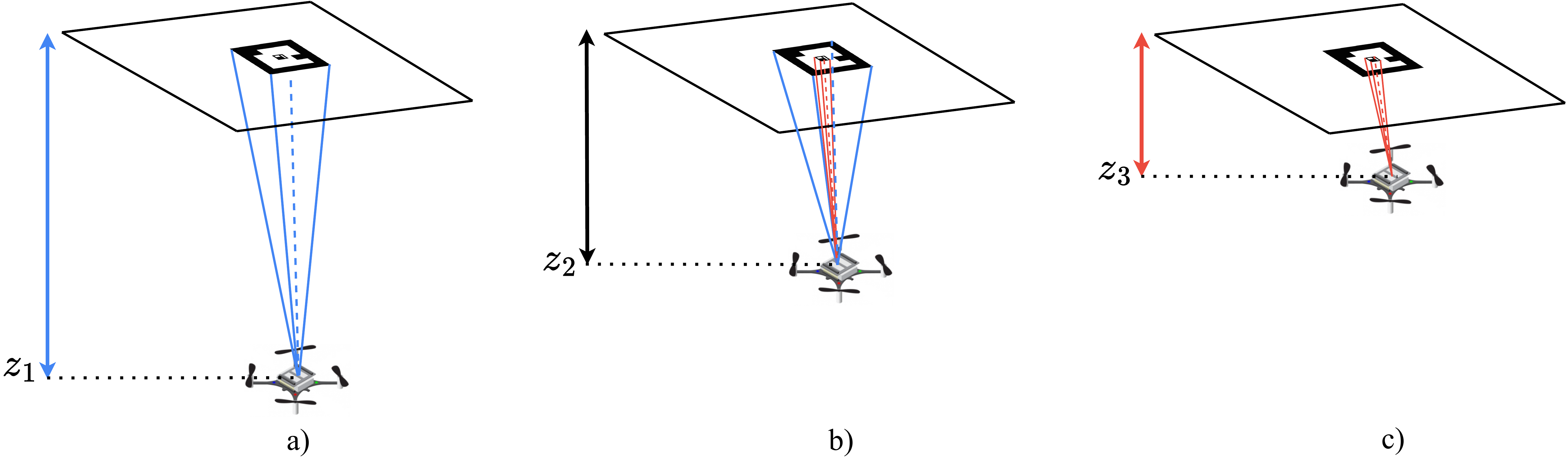}
\caption{Three stages of perching target pose estimation.}
\label{fig:3_stage_markers}
\end{figure*}
The data that we are looking to estimate and filter includes the relative $yaw$-angle $({}^{c}{{\psi }_{M}})$ and the translation ${}^{c}{{t}_{M}}$ between the body-fixed frame of the quadrotor and the reference frame of the detected marker, meaning that we need to define the state vector holding the elements: ${}^{c}{{P}_{M}}=\left\{ \psi ,\dot{\psi },{{t}_{x,}}{{{\dot{t}}}_{x,}}{{t}_{y,}}{{{\dot{t}}}_{y,}}{{t}_{z,}}{{{\dot{t}}}_{z,}} \right\}$, where $c,M$ is the camera and markers reference frame respectively. The operation of Kalman filter is illustrated in Fig. \ref{fig:kalman}. Having a state vector containing eight elements imply that we need a state transition vector initially defined as:
\begin{equation}
{{F}_{k}}=\left[ \begin{matrix}
   1 & \Delta t & 0 & 0 & 0 & 0 & 0 & 0  \\
   0 & 1 & 0 & 0 & 0 & 0 & 0 & 0  \\
   0 & 0 & 1 & \Delta t & 0 & 0 & 0 & 0  \\
   0 & 0 & 0 & 1 & 0 & 0 & 0 & 0  \\
   0 & 0 & 0 & 0 & 1 & \Delta t & 0 & 0  \\
   0 & 0 & 0 & 0 & 0 & 1 & 0 & 0  \\
   0 & 0 & 0 & 0 & 0 & 0 & 1 & \Delta t  \\
   0 & 0 & 0 & 0 & 0 & 0 & 0 & 1  \\
\end{matrix} \right]
\label{eqn:fk_matrix}
\end{equation}

where each non-zero element above the diagonal in each column of the matrix defines the time $\Delta t$ between the states. The measurement matrix, $H$ is then initiated as:
\begin{equation}
{{H}_{k}}=\left[ \begin{matrix}
   1 & 0 & 0 & 0 & 0 & 0 & 0 & 0  \\
   0 & 0 & 1 & 0 & 0 & 0 & 0 & 0  \\
   0 & 0 & 0 & 0 & 1 & 0 & 0 & 0  \\
   0 & 0 & 0 & 0 & 0 & 0 & 1 & 0  \\
\end{matrix} \right]
\label{eqn:hk_matrix}
\end{equation}

where a non-zero value represents the elements of which we want to measure and estimate. Thereafter, we define the state uncertainty matrix, $R_{k}={{k}_{1}}{{I}_{nxn}}$, where ${{k}_{1}}$ is the uncertainty factor and $n$ is the number of parameters that we want to estimate, in this case is four. Lastly, we define the process noise matrix, $Q_{k}={{k}_{2}}{{I}_{2nx2n}}$, where ${{k}_{2}}$ is a constant determining the magnitude of the process noise.

By defining the filter this way, we are assuming a constant velocity, which means we have not accounted for acceleration. This means that if the detected marker exhibits rapid movement in the image frame and detection is suddenly lost for several frames since the control signal to the drone relies on the estimated data given by the Kalman filter, it will follow a linear motion proportional to the estimated velocity. As a result, we reduced the speed for each state element exponentially when the desired marker was not detected for a number of frames in a row as bellow: 
\begin{equation}
{{{\hat{x}}'}_{{{k}_{n}}}}\left( {{v}_{k}} \right)={{{\hat{x}}'}_{{{k}_{n}}}}\left( {{v}_{k-1}} \right)\alpha 
\label{eqn:hk_matrix}
\end{equation}
where ${{k}_{n}}\le {{n}_{\max }}=8$ is denote a time step with an undetected marker, $\alpha =0.85$ denote the diminishing factor.

\subsection{Pose data merging}
Three stages of pose estimation is indicate in Fig. \ref{fig:3_stage_markers}. The relative pose of large marker $(M_{1})$ and the small marker $(M_{2})$ to the drone are defined as ${}^{c}{{P}_{{{M}_{1}}}}={{\left[ {}^{c}{{x}_{{{M}_{1}}}},{}^{c}{{y}_{{{M}_{1}}}},{}^{c}{{z}_{{{M}_{1}}}},{}^{c}{{\psi}_{{{M}_{1}}}} \right]}^{T}}$ and ${}^{c}{{P}_{{{M}_{2}}}}={{\left[ {}^{c}{{x}_{{{M}_{2}}}},{}^{c}{{y}_{{{M}_{2}}}},{}^{c}{{z}_{{{M}_{2}}}},{}^{c}{{\psi}_{{{M}_{2}}}} \right]}^{T}}$, respectively. Three stages are needed to calculate the relative estimated pose values of the drone to the target, ${{P}_{CF}}={{\left[ {{x}_{CF}},{{y}_{CF}},{{z}_{CF}},{{\psi }_{CF}} \right]}^{T}}$, is expressed as below.

\textbf{\textit{Stage 1:}} Only large marker $(M_{1})$ is detected (Fig. \ref{fig:3_stage_markers}a):
\begin{equation}
{{S}_{1}}=\left\{ ({{z}_{1}},{{z}_{2}},{{z}_{CF}})\in {{\mathbb{R}}^{3}}:{{z}_{2}}\le {{z}_{CF}}\le {{z}_{1}} \right\}
\label{eqn:s1}
\end{equation}

The pose of the drone to the perching target is calculated as
\begin{equation}
{{P}_{CF}}={}^{c}{{P}_{{{M}_{1}}}}
\label{eqn:s1_pose}
\end{equation}

\textbf{\textit{Stage 2:}} Both large marker $(M_{1})$ and small marker $(M_{2})$ are detected (Fig. \ref{fig:3_stage_markers}b):
\begin{equation}
{{S}_{2}}=\left\{ ({{z}_{2}},{{z}_{3}},{{z}_{CF}})\in {{\mathbb{R}}^{3}}:{{z}_{3}}\le {{z}_{CF}}\le {{z}_{2}} \right\}
\label{eqn:s2}
\end{equation}

Let define the pose vector of $(M_{1})$ and $(M_{2})$ as below:
\begin{equation}
{}^{c}{{P}_{{{M}_{12}}}}={{\left[ {}^{c}{{x}_{{{M}_{1}}}},{}^{c}{{x}_{{{M}_{2}}}},{}^{c}{{y}_{{{M}_{1}}}},{}^{c}{{y}_{{{M}_{2}}}},{}^{c}{{z}_{{{M}_{1}}}},{}^{c}{{z}_{{{M}_{2}}}},{}^{c}{{\psi }_{{{M}_{1}}}},{}^{c}{{\psi }_{{{M}_{2}}}} \right]}^{T}}
\label{eqn:pose_m12}
\end{equation}

We obtain the pose of the drone to the perching target as
\begin{equation}
{{P}_{CF}}=A{}^{c}{{P}_{{{M}_{12}}}}
\label{eqn:s2_pose}
\end{equation}

where
\begin{equation}
A=\left[ \begin{matrix}
   {{\sigma }_{x}} & 1-{{\sigma }_{x}} & 0 & 0 & 0 & 0 & 0 & 0  \\
   0 & 0 & {{\sigma }_{y}} & 1-{{\sigma }_{y}} & 0 & 0 & 0 & 0  \\
   0 & 0 & 0 & 0 & {{\sigma }_{z}} & 1-{{\sigma }_{z}} & 0 & 0  \\
   0 & 0 & 0 & 0 & 0 & 0 & {{\sigma }_{\psi }} & 1-{{\sigma }_{\psi }}  \\
\end{matrix} \right]
\label{eqn:A}
\end{equation}

${{\sigma }_{x}},{{\sigma }_{y}},{{\sigma }_{z}},{{\sigma }_{\Psi }}$ is chosen by applying Least Mean Square (LMS) \cite{b22} filtering algorithm that minimizes cost function between the actual pose and the estimated pose of large marker $(M_{1})$ and small marker $(M_{2})$ from the camera:
\begin{equation}
C(n)=E\{|e(n){{|}^{2}}\}
\label{eqn:lms}
\end{equation}

where $e(n)$ is the error at the current state $n$ and $E\{\cdot\}$ denotes the expected value.

\textbf{\textit{Stage 3:}} Only small marker $(M_{2})$ is detected (Fig. \ref{fig:3_stage_markers}c):
\begin{equation}
{{S}_{3}}=\left\{ ({{z}_{3}},{{z}_{CF}})\in {{\mathbb{R}}^{3}}:{{z}_{CF}}\le {{z}_{3}} \right\}
\label{eqn:s3}
\end{equation}

The pose of the drone to the perching target is determined as:
\begin{equation}
{{P}_{CF}}={}^{c}{{P}_{{{M}_{2}}}}
\label{eqn:s3_pose}
\end{equation}
\section{Experiments}
\subsection{Testbed}

\begin{figure}[t!]
\centering
\includegraphics[width=8cm]{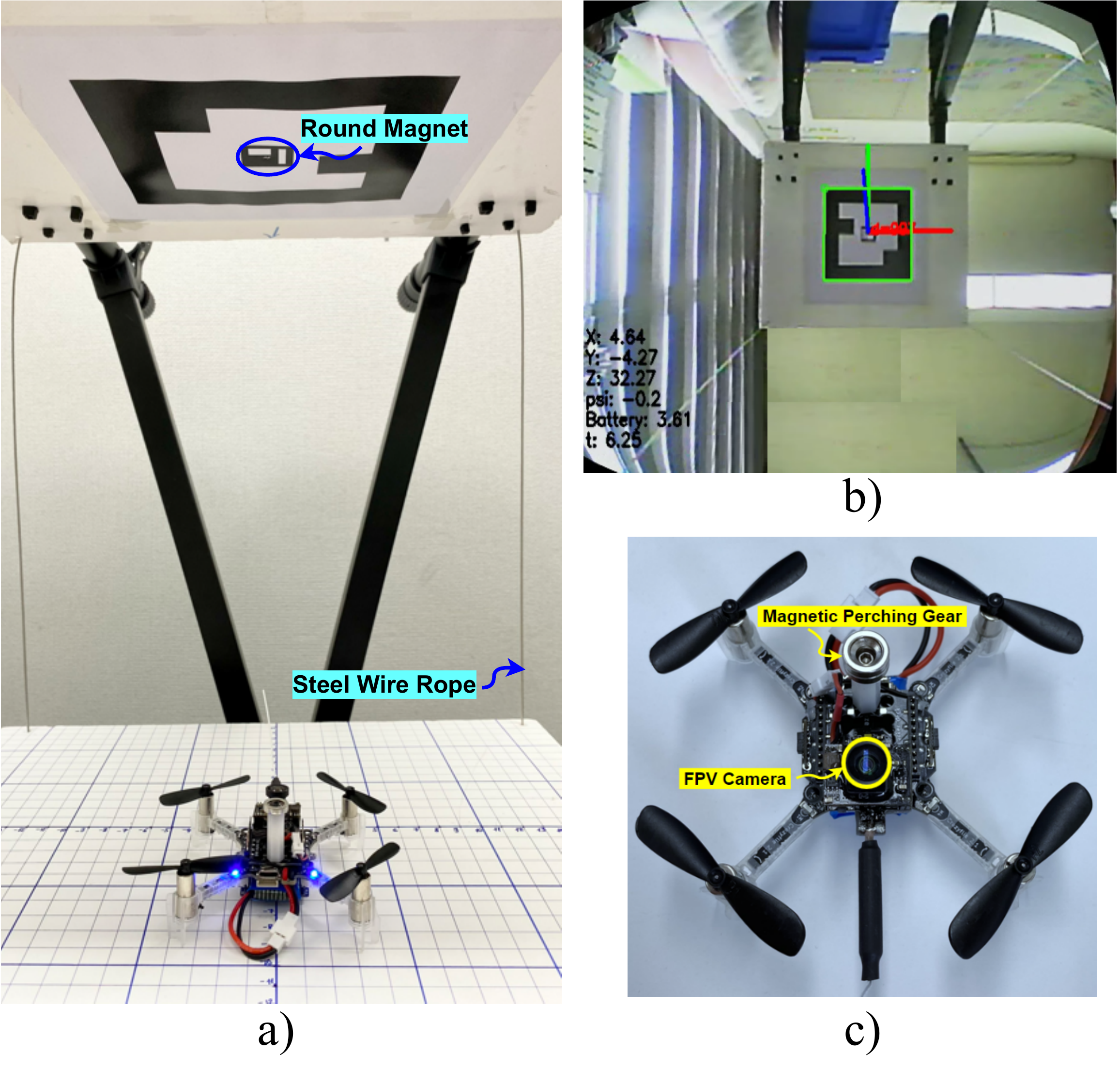}
\caption{Experimental testbed.}
\centering
\label{fig:testbed}
\end{figure}

Figure \ref{fig:testbed}a) shows the testbed for evaluating the proposed estimator. It consists of 2 parts: a perching plate and a lifting plate. Their four corners are connected by steel wire ropes. On the perching plane, the designed marker is printed and placed downward. A round magnet with a diameter of $25mm$ is installed in the center of the perching target of multi-markers. The lifting plate is plotted in vertical and horizontal lines along the x and y-axis with one-centimeter spacing. Using this platform, we can adjust and measure the actual values for reference in $x$, $y$, $z$ directions.

The drone used in this research is a Crazyflie 2.1 nano UAV as shown in Fig. \ref{fig:testbed}c). The FPV camera is mounted on drone and pointed upwards in eye-in-hand configuration for detecting visual markers and estimating relative pose. Image frame collected from the camera have a size of 640 x 480 pixels. A small and strong magnetic perching gear is held on the top of the drone. The total weight of constructed Crazyflie is only 39 grams, including the battery.

Figure \ref{fig:testbed}b) shows the drone's camera view while it is perching. There are red, green, and blue lines illustrating the $z$, $x$, and $y$-axis, respectively, with the green square enclosing the detected marker. The left-down corner shows the estimated values of the target pose in centimeters of $x$, $y$, $z$, and degrees of $yaw$-angle.

\begin{figure}[t!]
\centering
\includegraphics[width=8cm]{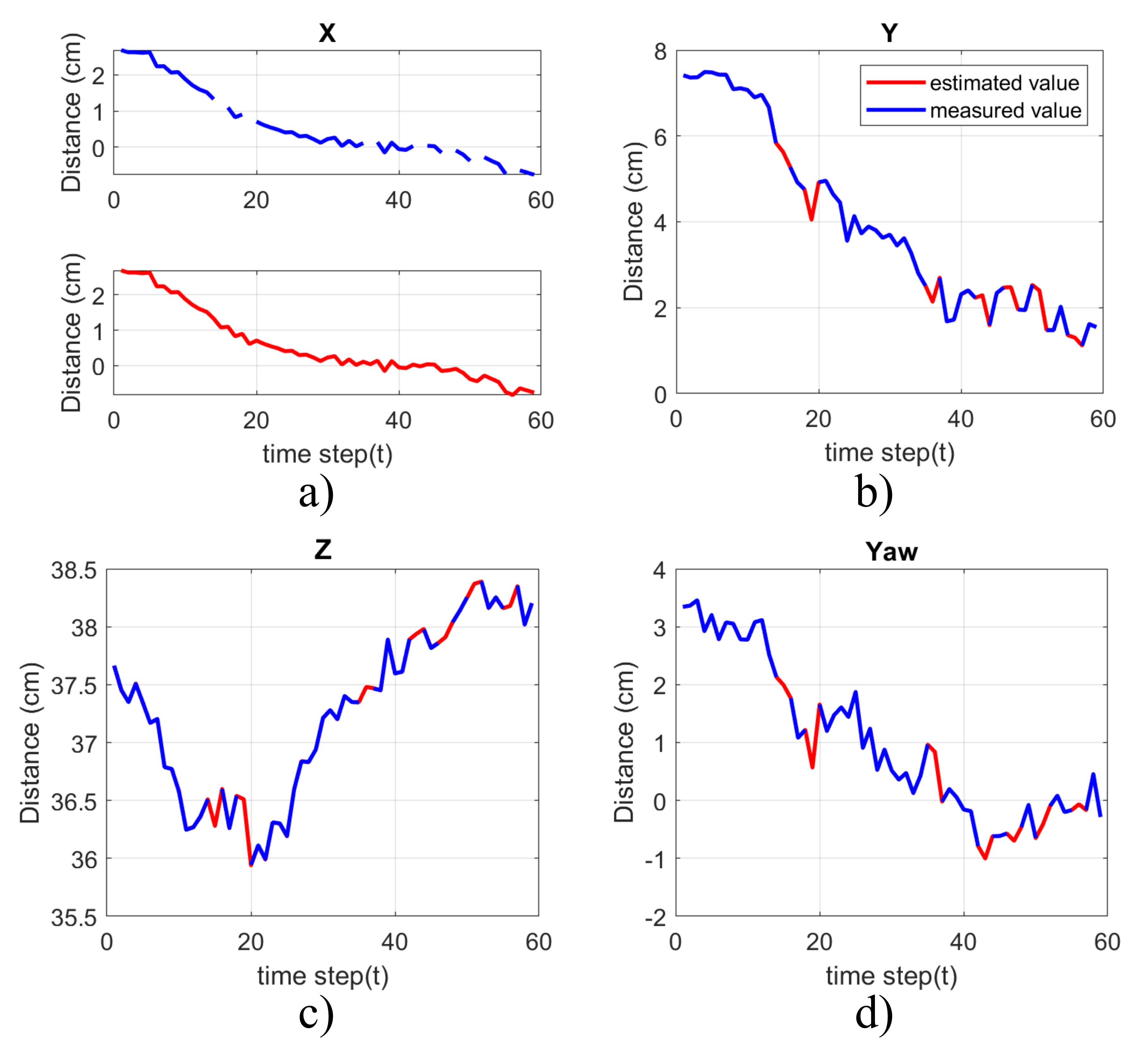}
\caption{The target pose data after utilizing the Kalman filter.}
\centering
\label{fig:kalman_results}
\end{figure}

\subsection{Experimental results}

Our experiments showed that the large $(M_{1})$ and small $(M_{2})$ markers begin to be detected at a maximum distance $z_{1}=115$cm, $z_{2}=25$cm, based on the experimental measurements performed. In the absence of a large marker, the minimum relative distance is $z_{3}=15$cm.

Figure \ref{fig:kalman_results} shows plots over a short tracking sequence where the drone was moved in several different positions. This blue line represents the estimated values obtained after using the Kalman filter, whereas the red line shows the raw data output from the ArUco detection and pose estimation function of OpenCV. In the real cases of perching target pose calculation, the output data is not consecutive without the Kalman filter estimation. Drone crashes will result from this. As can be seen upon inspection of the figure, during the measurement period, the Kalman filter is capable of making a good estimate of the movement when the measurement input data is lost. Especially at $t\approx18$s, no marker is detected for approximately one second, and Kalman filters predict marker movement using velocity at this time, decreasing the velocity by the factor $\alpha$ for each iteration until the maximum number of frames $n_{max}$ is reached.

To determine the values of the merging coefficient, the estimated values and actual values of them were collected at ten random positions and directions. There are $100$ consecutive data values at each point. The LMS algorithm is then used to find the optimal coefficient values with the least amount of error between actual and estimated data. We achieve ${{\sigma}_{x}}=0.275$, ${{\sigma}_{y}}=0.306$, ${{\sigma}_{z}}=0.728$, ${{\sigma}_{\psi}}=0.469$.

Table \ref{tb1} shows relative position data $(x_{CF},y_{CF},z_{CF})$ including the actual values, estimated values and the peak-to-peak magnitudes of the error between them in centimeters. We measured and estimated the position values at five different point of the drone, the peak-to-peak estimated position error magnitudes in the range of 0.6cm and decreasing when the drone approaches near and close to the center of the target.

Similarly, we also gathered the relative heading angle in degree $(\psi_{CF})$. There is no significant difference between actual values and estimated values as shown in Table \ref{tb2}. When the $yaw$-angle is close to the straight forward direction, the rounded error is become zero.

\begin{table}[!b]
\centering
\caption{The estimation of relative position values in centimeter.}\label{tb1}
\resizebox{\columnwidth}{!}{%
\begin{tabular}{|c|c|c|c|c|c|}
\hline
\textbf{\begin{tabular}[c]{@{}c@{}}Actual \\ Position\end{tabular}}    & (0, 0, 24)        & (3, 5, 16)       & (5,8,11)        & (10,12,7)       & (6,-5,18)        \\ \hline
\textbf{\begin{tabular}[c]{@{}c@{}}Estimated \\ Position\end{tabular}} & (0.3, -0.3, 24.4) & (2.6, 5.3, 16.4) & (5.4,8.4, 11.3) & (10.5,12.6,7.2) & (6.3,-5.4,18.5)  \\ \hline
\textbf{\begin{tabular}[c]{@{}c@{}}Est. Position\\ Error\end{tabular}} & (0.3, -0.3, 0.4)  & (0.4, 0.3, 0.4)  & (0.4, 0.4, 0.3) & (0.5, 0.6, 0.2) & (0.3, -0.4, 0.5) \\ \hline
\end{tabular}%
}
\end{table}

\begin{table}[!b]
\centering
\caption{The estimation of relative heading angle values in degree.}\label{tb2}
\resizebox{0.8\columnwidth}{!}{%
\begin{tabular}{|
>{}c |m{0.7cm}|m{0.7cm}|m{0.7cm}|m{0.7cm}|m{0.7cm}|m{0.7cm}|m{0.7cm}|m{0.7cm}|}
\hline
\textbf{\begin{tabular}[c]{@{}c@{}}Actual \\ Heading Angle\end{tabular}}    & \hfil175 & \hfil-160 & \hfil135 & \hfil90 & \hfil30 & \hfil10 & \hfil5 & \hfil0 \\ \hline
\textbf{\begin{tabular}[c]{@{}c@{}}Estimated \\ Heading Angle\end{tabular}} & \hfil178 & \hfil-157 & \hfil137 & \hfil92 & \hfil31 & \hfil9  & \hfil5 & \hfil1 \\ \hline
\textbf{\begin{tabular}[c]{@{}c@{}}Est. Heading Angle\\ Error\end{tabular}} & \hfil3   & \hfil-3   & \hfil2   & \hfil2  & \hfil1  & \hfil1  & \hfil0 & \hfil0 \\ \hline
\end{tabular}%
}
\end{table}

\section{Conclusion}

This research proposed a vision-based target pose estimation method using multiple markers for high-precision nano drone perching at both wide and close ranges. The multi-marker was designed as a perching target, and the image frame was captured using a monocular camera mounted on Crazyflie. After that, the markers will be detected and the pose calculated. Kalman filters were applied to fill in the empty gaps caused by missing transmissions or detection failures. Then the pose from multiple markers was combined with the optimal coefficients achieved from the Least Mean Squared algorithm. Finally, a testbed was created to validate the real-world measurement of relative distances and heading angles between the vehicle and target. Experimental results have confirmed the efficiency and practicality of the presented approach. With this system, it is possible to estimate the target pose for perching at the millimeter level of accuracy. This technique can be utilized for nano UAVs, multirotors, and space vehicles in the perching and docking tasks. 

There are still directions for further development to achieve even more remarkable results. In future works, the proposed target pose estimator will be used for conducting the perching of nano drones in disturbance environments.


\section*{Acknowledgments}\label{Acknowledgments}

This work was supported by Future Space Navigation \& Satellite Research Center through the National Research Foundation funded by the Ministry of Science and ICT, the Republic of Korea (2022M1A3C2074404).

This research was supported by the MSIT (Ministry of Science and ICT), Korea, under the ITRC (Information Technology Research Center) support program (IITP-2022-2018-0-01423) supervised by the IITP (Institute for Information \& Communications Technology Planning \& Evaluation).

\AEnameShow
\end{document}